\documentclass[conference]{IEEEtran}
\IEEEoverridecommandlockouts
\usepackage{cite}
\usepackage{amsmath,amssymb,amsfonts}
\usepackage{graphicx}
\usepackage{textcomp}
\usepackage{xcolor}
\usepackage{balance}
\usepackage{cite}
\usepackage{hyperref}
\usepackage{amsmath,amssymb,amsfonts}
\usepackage{amsmath}
\usepackage{amsthm}

\usepackage{graphicx}
\usepackage{textcomp}
\usepackage{xcolor}
\usepackage{graphicx}
\usepackage{float}
\usepackage{subfigure}
\usepackage{amsmath}
\usepackage{amsfonts,amssymb}
\usepackage{mathrsfs}
\usepackage{mathtools}
\usepackage{bm}
\usepackage{multirow}
\usepackage{array}
\usepackage{amssymb}
\usepackage{amsmath}
\usepackage{cite}
\usepackage{url}
\usepackage{xcolor}
\usepackage{cite,graphicx,amsmath,amssymb}
\usepackage{subfigure}
\usepackage{fancyhdr}
\usepackage{mdwmath}
\usepackage{mdwtab}
\usepackage{caption}
\usepackage{amsthm}
\usepackage{setspace}
\usepackage{bm}
\usepackage{mathtools}
\usepackage{dsfont}
\usepackage{bbm}
\usepackage{algorithm}
\usepackage{algorithmic}

\newtheorem{theorem}{Theorem}

\newtheorem{lemma}{Lemma}

\newtheorem{corollary}{Corollary}

\makeatletter
\newcommand{\biggg}{\bBigg@{3}}
\newcommand{\Biggg}{\bBigg@{3.5}}
\makeatother
\def\BibTeX{{\rm B\kern-.05em{\sc i\kern-.025em b}\kern-.08em
    T\kern-.1667em\lower.7ex\hbox{E}\kern-.125emX}}
\begin{document}
\title{Movable Antenna Aided Physical Layer Security with No Eavesdropper CSI}
\author{Zhenqiao Cheng, Chongjun Ouyang, and Xingqi Zhang
\thanks{Z. Cheng, N. Li, J. Zhu, X. She, and P. Chen are with the 6G Research Centre, China Telecom Beijing Research Institute, Beijing, 102209, China (e-mail: \{chengzq, linanxi, zhujc, shexm, chenpeng11\}@chinatelecom.cn).}
\thanks{B. Ning is with the National Key Laboratory of Science and Technology on Communications, University of Electronic Science and Technology of China, Chengdu 611731, China (e-mail: boydning@outlook.com).}
\thanks{C. Ouyang is with the School of Electrical and Electronic Engineering, University College Dublin, Dublin, D04 V1W8, Ireland (e-mail: chongjun.ouyang@ucd.ie).}}
\author{\IEEEauthorblockN{Zhenqiao Cheng$^{\star}$, Chongjun Ouyang$^{\ddag}$, and Xingqi Zhang$^{\dag}$}
$^{\star}$6G Research Centre, China Telecom Beijing Research Institute, Beijing, 102209, China\\
$^\ddag$University College Dublin, Dublin, Ireland \& Queen Mary University of London, London, U.K.\\
$^\dag$Department of Electrical and Computer Engineering, University of Alberta, Canada}
\maketitle
\begin{abstract}
A novel movable antenna (MA)-aided secure transmission framework is proposed to enhance the secrecy transmission rate without relying on the eavesdropper's channel state information. Within this framework, a joint beamforming and jamming scheme is proposed, where the power of the confidential signal is minimized by optimizing the positions of the MAs, and the residual power is used to jam the eavesdropper. An efficient gradient-based method is employed to solve this non-convex problem. Numerical results are provided to demonstrate the superiority of the MA-based framework over systems using traditional fixed-position antennas in secure transmission.
\end{abstract}
\begin{IEEEkeywords}
Antenna position, movable antenna, physical layer security, secrecy rate.
\end{IEEEkeywords}
\section{Introduction}
In traditional multiple-antenna systems, antennas remain stationary, limiting their ability to fully exploit spatial variations within a given transmit/receive region \cite{Liu2024,Liu2023}. This limitation becomes particularly evident when using a limited number of antennas. To address this, the concept of movable antennas (MAs), or fluid antennas, has been introduced \cite{Zhu2024,Zhu2024his}. MAs are designed to overcome the challenges posed by conventional fixed-position antennas (FPAs). By connecting to radio frequency (RF) chains via flexible cables and incorporating real-time position adjustment mechanisms, MAs gain the flexibility to dynamically adjust their \emph{positions}. This adaptability enables MAs to reshape the wireless channel, resulting in enhanced wireless transmission capabilities \cite{New2024}.

Recently, integrating MAs to enhance physical layer security \cite{Chen2017} has gained significant attention in research. The motivation behind combining MAs with physical layer security is to leverage the additional spatial degrees of freedom (DoFs) provided by MAs to further improve the secrecy transmission rate. Several numerical studies have explored this concept; see \cite{Cheng2023_1,Hu2024,Tang2024,Hu2024movable,Feng2024} and relevant references for further details. However, most existing works are based on the ideal assumption that the eavesdropper's channel state information (CSI) is perfectly known; see \cite{Cheng2023_1,Hu2024,Tang2024}. This assumption is impractical, as eavesdroppers are typically hidden and passive, and do not actively exchange CSI with the transmitter. It is worth noting that some works employ partial eavesdropper CSI, such as statistical CSI \cite{Hu2024movable} or location information \cite{Feng2024}, to optimize the positions of the MAs. However, in cases where the eavesdropper is fully hidden or silent, obtaining this information is also challenging. Consequently, the secrecy schemes proposed in these existing works have limited applicability in practical scenarios.

Motivated by this, we study MA-aided secrecy transmission without the eavesdropper's CSI in this paper. The main contributions are summarized as follows: {\romannumeral1}) We propose a joint beamforming and jamming scheme to enhance the secrecy performance of an MA-aided wiretap channel. In this scheme, we minimize the power of the confidential signal to meet the quality of service (QoS) at the legitimate user while allocating all residual power to transmit artificial noise (AN) to jam the eavesdropper. {\romannumeral2}) We propose an efficient gradient-based algorithm to optimize the positions of the MAs, addressing the non-convex optimization problem. {\romannumeral3}) To provide further insights, we derive a novel expression for the achievable secrecy rate and discuss the case when the transmitter employs a linear array. {\romannumeral4}) Numerical results demonstrate that the proposed MA-based transmission offers more DoFs for improving the secrecy rate compared to conventional FPA-based systems, even in the absence of eavesdropper's CSI.

\begin{figure}[!t]
\centering
    \subfigbottomskip=-5pt
	\subfigcapskip=-5pt
\setlength{\abovecaptionskip}{0pt}
\includegraphics[height=0.18\textwidth]{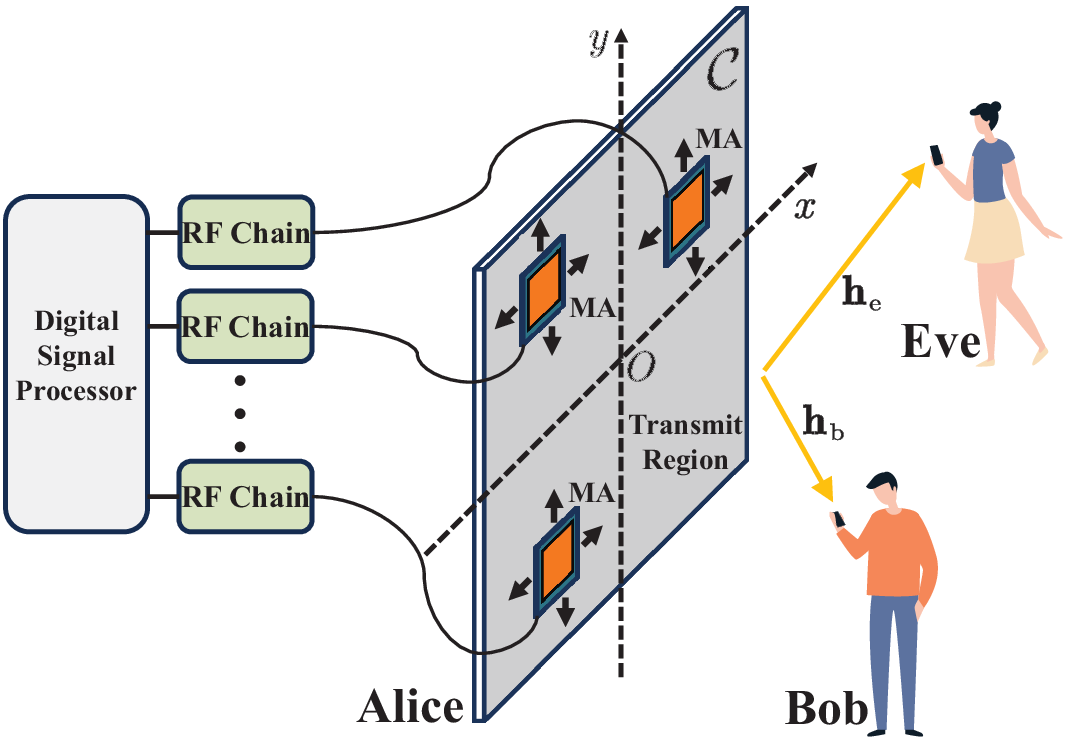}
\caption{An MA-aided secure communication system.}
\label{System_Model_PHY}
\vspace{-10pt}
\end{figure}

\section{System Model}
\subsection{Secure Transmission Model}
Consider a secure communication system aided by MAs, comprising Alice, Bob, and Eve, as illustrated in {\figurename} {\ref{System_Model_PHY}}. In this scenario, Alice is equipped with $N$ transmit MAs, while both Bob and Eve are equipped with a single receive FPA. The MAs, implemented using liquid-metal-based antennas, can adjust their positions. The position of the $n$th MA is represented by the Cartesian coordinates ${\mathbf{t}}_n=[x_n, y_n]^{\mathsf{T}}\in{\mathcal{C}}$ for $n\in{\mathcal{N}}\triangleq\{1,\ldots,N\}$, where $\mathcal{C}$ denotes the specified two-dimensional transmit region within which the MAs can move freely. Without loss of generality, we assume $\mathcal{C}$ to be a square region with dimensions $A\times A$, as depicted in {\figurename} {\ref{System_Model_PHY}}.

We consider quasi-static block-fading channels and focus on a specific fading block, where the multi-path channel components at any location within $\mathcal{C}$ are assumed to be fixed. In the absence of Eve's CSI, we propose a joint beamforming and jamming scheme in which Alice simultaneously transmits both information and AN signals. The received signals at Bob and Eve are expressed as follows:
{\setlength\abovedisplayskip{2pt}
\setlength\belowdisplayskip{2pt}
\begin{align}
y_{i}={\mathbf{h}}_{i}^{\mathsf{H}}({\mathbf{w}}s+{\mathbf{z}})+n_{i},~i\in\{{\rm{b}},{\rm{e}}\},
\end{align}
}where the subscripts ``${\rm{b}}$'' and ``${\rm{e}}$'' refer to Bob and Eve, respectively. Additionally, ${\mathbf{w}}\in{\mathbbmss{C}}^{N\times1}$ is the transmit beamforming vector, ${\mathbf{z}}\in{\mathbbmss{C}}^{N\times1}$ is the AN vector, and ${\mathbf{h}}_{i}\in{\mathbbmss{C}}^{N\times1}$ represents the channel vector from Alice to node $i$. The term $n_i\sim{\mathcal{CN}}(0,\sigma_i^2)$ denotes the additive complex white Gaussian noise at node $i$, with $\sigma_i^2$ representing the noise power. The transmitted signal $s\in{\mathbbmss{C}}$ is assumed to be Gaussian with zero mean and unit variance, i.e., ${\mathbbmss{E}}\{s\}=0$ and ${\mathbbmss{E}}\{\lvert s\rvert^2\}=1$. The signal-to-noise ratio (SNR) at node $i$ is given by 
{\setlength\abovedisplayskip{2pt}
\setlength\belowdisplayskip{2pt}
\begin{align}
\gamma_{i}=\frac{\lvert{\mathbf{h}}_{i}^{\mathsf{H}}{\mathbf{w}}\rvert^2}{\sigma_i^2+{\mathbf{h}}_{i}^{\mathsf{H}}{\mathbf{C}}{\mathbf{h}}_{i}},~i\in\{{\rm{b}},{\rm{e}}\},
\end{align}
}where ${\mathbf{C}}={\mathbbmss{E}}\{{\mathbf{z}}{\mathbf{z}}^{\mathsf{H}}\}\in{\mathbbmss{C}}^{N\times N}$ is the covariance matrix of the AN. It is assumed that the AN is located in the null space of $\mathbf{h}_{\rm{b}}$, which implies ${\mathbf{h}}_{\rm{b}}^{\mathsf{H}}{\mathbf{C}}{\mathbf{h}}_{{\rm{b}}}=0$. Therefore, the received SNR at Bob is given by $\gamma_{\rm{b}}=\frac{\lvert{\mathbf{h}}_{\rm{b}}^{\mathsf{H}}{\mathbf{w}}\rvert^2}{\sigma_{\rm{b}}^2}$. The achievable secrecy rate for this scheme is expressed as follows:
{\setlength\abovedisplayskip{2pt}
\setlength\belowdisplayskip{2pt}
\begin{align}\label{Secrecy_Rate}
{\mathcal{R}}_{\rm{s}}=\max\{\log_2(1+\gamma_{\rm{b}})-\log_2(1+\gamma_{\rm{e}}),0\}.
\end{align}
}Furthermore, Alice is subject to a power constraint $\lVert{\mathbf{w}}\rVert^2+{\mathsf{tr}}({\mathbf{C}})\leq P$, where $P$ represents the total transmit power.
\subsection{Channel Model}
We consider the field-response based channel model \cite{Ma2023}, as expressed by ${\mathbf h}_{i}=[h_{i}({\mathbf{t}}_1),\ldots, h_{i}({\mathbf{t}}_N)]^{\mathsf{T}}$, where 
{\setlength\abovedisplayskip{2pt}
\setlength\belowdisplayskip{2pt}
\begin{align}\label{Channel_Model}
h_{i}({\mathbf{t}})\triangleq\sum\nolimits_{\ell=1}^{L_i}\sqrt{\frac{\mu_i}{L_i}}\sigma_{i,\ell}{\rm{e}}^{-{\rm{j}}\frac{2\pi}{\lambda}{\mathbf{t}}^{\mathsf{T}}{\bm\rho}_{i,\ell}},~i\in\{{\rm{b}},{\rm{e}}\},
\end{align}
}and ${\bm\rho}_{i,\ell}=[\sin{\theta_{i,\ell}}\cos{\phi_{i,\ell}},\cos{\theta_{i,\ell}}]^{\mathsf{T}}$. Here, $L_i$ represents the number of channel paths, $\theta_{i,\ell}$ and $\phi_{i,\ell}$ are the elevation and azimuth angles of the $\ell$th path. Additionally, $\mu_i$ denotes the path loss, $\sigma_{i,\ell}\sim{\mathcal{CN}}(0,1)$ represents the small-scale fading, and $\lambda$ is the wavelength. 
\section{Secure Transmission Without Eve's CSI}
In contrast to previous studies \cite{Cheng2023_1,Hu2024,Tang2024}, which assume that Alice has access to the CSI of both Bob and Eve, we address a more realistic scenario. In this case, Alice has knowledge of Bob's CSI, while Eve's CSI remains entirely unknown. Given this premise, directly maximizing the secrecy rate \eqref{Secrecy_Rate} is not feasible. To improve security under these conditions, a potential approach involves increasing the information rate at Bob while minimizing information leakage to Eve.

Building on the insights from \cite{Wang2012,Wang2020}, our proposed joint beamforming and jamming scheme consists of two key steps. \emph{Firstly}, the positions of the MAs and the transmit beamformer are jointly designed to meet the specified quality of service (QoS) target for Bob. This ensures that the remaining power, $P-\lVert{\mathbf{w}}\rVert^2$, is maximized for generating AN to interfere with Eve. \emph{Secondly}, the AN is designed to satisfy ${\mathbf{z}}\perp{\mathbf{h}}_{\rm{b}}$, ensuring that it is projected onto the null space of Bob's spatial channel, thereby exclusively targeting Eve. Both steps are optimized to maximize the secrecy rate, even in the absence of Eve's CSI.

Our primary objective in the first step is to minimize $\lVert{\mathbf{w}}\rVert^2$ by optimizing both $\mathbf{w}$ and $\{{\mathbf{t}}_n\}_{n=1}^{N}$, subject to the QoS constraint at Bob. To simplify and derive a performance upper bound, we assume that the MA positions $\{{\mathbf{t}}_n\}_{n=1}^{N}$ are optimized using Bob's full instantaneous CSI, which includes both the angle information $\{{\bm\rho}_{{\rm{b}},\ell}\}_{\ell=1}^{L_{\rm{b}}}$ and the small-scale fading coefficients $\{\sigma_{\rm{b},\ell}\}_{\ell=1}^{L_{\rm{b}}}$. The procedure for obtaining these coefficients involves several channel estimation steps. First, conventional CSI acquisition methods are applied to obtain the channel vector ${\mathbf{h}}_{\rm{b}}$ for a given position set $\{{\mathbf{t}}_n\}_{n=1}^{N}$. Then, the angle and reflection coefficients for each path, i.e., $\{{\bm\rho}_{{\rm{b}},\ell},\sigma_{\rm{b},\ell}\}_{\ell=1}^{L_{\rm{b}}}$, can be extracted from the estimate of ${\mathbf{h}}_{\rm{b}}$ using the MUSIC and APES algorithms \cite{Li1996,Schmidt1986}. 

These considerations imply that ensuring QoS at Bob involves guaranteeing his instantaneous received SNR. Mathematically, the problem of joint transmit beamforming and antenna positioning can be formulated as follows:
{\setlength\abovedisplayskip{2pt}
\setlength\belowdisplayskip{2pt}
\begin{subequations}\label{Power_Min_Problem_Instantaneous}
\begin{align}
\min\nolimits_{{\mathbf{T}},{\mathbf{w}}}~&\lVert{\mathbf{w}}\rVert^2\label{Power_Min_Problem_Instantaneous_Obj}\\
{\rm{s.t.}}~&\lvert{\mathbf{h}}_{\rm{b}}^{\mathsf{H}}{\mathbf{w}}\rvert^2/\sigma_{\rm{b}}^2\geq \gamma,\label{Power_Min_Problem_Instantaneous_Cons1}\\
&{\mathbf{t}}_n\in{\mathcal{C}},\forall n\in{\mathcal{N}},\lVert {\mathbf{t}}_n-{\mathbf{t}}_{n'}\rVert\geq D,n\ne n',\label{Power_Min_Problem_Instantaneous_Cons2}
\end{align}
\end{subequations}
}where ${\mathbf{T}}=[{\mathbf{t}}_1\ldots{\mathbf{t}}_N]\in{\mathbbmss{R}}^{2\times N}$, $\gamma>0$ represents Bob's target SNR, and $D$ is the minimum distance required between each pair of antennas to avoid mutual coupling effects. Once the optimized $\{{\mathbf{t}}_n\}_{n=1}^{N}$ is obtained, the AN can be designed based on the instantaneous knowledge of $\mathbf{h}_{\rm{b}}$.

Unlike FPA-based secure communication systems, the transmit beamforming vector in MA-based systems is dependent on the positions of the MAs. The problem defined in \eqref{Power_Min_Problem_Instantaneous} poses significant challenges due to its non-convex constraints and tightly coupled variables. Therefore, we aim to propose an efficient algorithms to address these challenges.
\section{Joint Beamforming and Jamming Design}
\subsection{Transmit Beamforming and Antenna Positioning Design}
For a given MA position $\mathbf{T}$, it is known that maximum ratio transmission (MRT) provides the optimal transmit beamforming solution to problem \eqref{Power_Min_Problem_Instantaneous}. Specifically, the optimal beamforming vector is ${\mathbf{w}}^{\star}=\sqrt{P_{\rm{T}}}\frac{{\mathbf{h}}_{\rm{b}}}{\lVert{\mathbf{h}}_{\rm{b}}\rVert}$, where $P_{\rm{T}}$ represents the transmit power for the beamformer. Substituting ${\mathbf{w}}^{\star}$ into problem \eqref{Power_Min_Problem_Instantaneous} transforms the problem into the following:
{\setlength\abovedisplayskip{2pt}
\setlength\belowdisplayskip{2pt}
\begin{subequations}\label{Power_Min_Problem1}
\begin{align}
\min\nolimits_{{\mathbf{T}},P_{\rm{T}}}~&P_{\rm{T}}\\
{\rm{s.t.}}~&{P_{\rm{T}}}/{\sigma_{\rm{b}}^2}\lVert{\mathbf{h}}_{\rm{b}}\rVert^2 \geq \gamma,~\eqref{Power_Min_Problem_Instantaneous_Cons2}.
\end{align}
\end{subequations}
}It can be verified that the optimal power is given by $P_{\rm{T}}^{\star}=\frac{\gamma{\sigma_{\rm{b}}^2}}{\lVert{\mathbf{h}}_{\rm{b}}\rVert^2}$. Therefore, minimizing the power is equivalent to maximizing Bob's channel power gain, i.e.,
{\setlength\abovedisplayskip{2pt}
\setlength\belowdisplayskip{2pt}
\begin{align}\label{Power_Min_Problem2}
\max\nolimits_{{\mathbf{T}}}~\lVert{\mathbf{h}}_{\rm{b}}\rVert^2=\sum\nolimits_{n=1}^{N}\lvert h_{\rm{b}}({\mathbf{t}}_n)\rvert^2\quad{\rm{s.t.}}~\eqref{Power_Min_Problem_Instantaneous_Cons2}.
\end{align}
}

Problem \eqref{Power_Min_Problem2} is a single-variable optimization problem, significantly simpler than problem \eqref{Power_Min_Problem_Instantaneous}. However, it remains challenging due to non-convex constraints and the tight coupling of $\{{\mathbf{t}}_n\}_{n=1}^{N}$. To address this, we partition the variable set into $N$ distinct blocks $\{\mathbf{t}_n\}_{n=1}^{N}$. We then solve $N$ subproblems, each optimizing $\mathbf{t}_n$ for $n=1,\ldots,N$, while keeping all other variables fixed. This developed algorithm iteratively solves these subproblems in an alternating manner, providing a locally optimal solution for problem \eqref{Power_Min_Problem_Instantaneous}.

Given $\{{\mathbf{t}}_{n'}\}_{n'\ne n}$, the marginal problem for ${\mathbf{t}}_n$ is formulated as follows:
{\setlength\abovedisplayskip{2pt}
\setlength\belowdisplayskip{2pt}
\begin{align}\label{Power_Min_Problem3}
\max\nolimits_{{\mathbf{t}}_n\in{\mathcal{S}}_n}~f_n({\mathbf{t}}_n)&\triangleq\lvert h_{\rm{b}}({\mathbf{t}}_n)\rvert^2\\
&=\frac{\mu_{\rm{b}}}{L_{\rm{b}}}\left\lvert\sum\nolimits_{\ell=1}^{L_{\rm{b}}}\sigma_{{\rm{b}},\ell}{\rm{e}}^{-{\rm{j}}\frac{2\pi}{\lambda}{\mathbf{t}}_n^{\mathsf{T}}{\bm\rho}_{{\rm{b}},\ell}}\right\rvert^2,
\label{Objective_Function}
\end{align}
}where ${\mathcal{S}}_n\triangleq\{{\mathbf{x}}|{\mathbf{x}}\in{\mathcal{C}},\lVert {\mathbf{x}}-{\mathbf{t}}_{n'}\rVert\geq D,\forall n\ne n'\}$ represents the candidate location set for the $n$th MA. Due to the intractability of $f_n(\cdot)$, we employ the gradient descent method with backtracking line search to find stationary points for subproblem \eqref{Power_Min_Problem3} \cite{Boyd2004}. For clarity, we rewrite \eqref{Objective_Function} as follows:
{\setlength\abovedisplayskip{2pt}
\setlength\belowdisplayskip{2pt}
\begin{equation}
\begin{split}
f_n({\mathbf{t}}_n)&=\lvert h_{\rm{b}}({\mathbf{t}}_n)\rvert^2=h_{\rm{b}}({\mathbf{t}}_n)h_{\rm{b}}^{*}({\mathbf{t}}_n)=\frac{\mu_{\rm{b}}}{L_{\rm{b}}}\lvert\sigma_{{\rm{b}},\ell}\rvert^2\\
&+\frac{\mu_{\rm{b}}}{L_{\rm{b}}}\sum_{\ell=1}^{L_{\rm{b}}}\sum_{\ell'\ne \ell}\lvert\sigma_{{\rm{b}},\ell}\sigma_{{\rm{b}},\ell'}\rvert
{\rm{e}}^{-{\rm{j}}(\frac{2\pi}{\lambda}
{\mathbf{t}}_n^{\mathsf{T}}{\bm\rho}_{\rm{b}}^{\ell,\ell'}+\theta_{\rm{b}}^{\ell,\ell'})},
\end{split}
\end{equation}
}where ${\bm\rho}_{\rm{b}}^{\ell,\ell'}={\bm\rho}_{{\rm{b}},\ell}-{\bm\rho}_{{\rm{b}},\ell'}$ and $\theta_{\rm{b}}^{\ell,\ell'}=\angle{\sigma}_{\ell',{\rm{b}}}-\angle{\sigma}_{\ell,{\rm{b}}}$. Using Euler's formula and the principles of complex-valued matrix differentiation \cite{Gesbert2007}, we calculate the gradient of $f_n({\mathbf{t}}_n)$ with respect to ${\mathbf{t}}_n$ as follows \cite{Cheng2024cl,Cheng2023air}:
{\setlength\abovedisplayskip{2pt}
\setlength\belowdisplayskip{2pt}
\begin{equation}\label{Der_MA}
\begin{split}
\nabla_{{\mathbf{t}}_n}f_{n}=
\sum_{\ell=1}^{L_{\rm{b}}}\sum_{\ell'\ne \ell}\frac{4\pi \mu_{{\rm{b}}}\sin(\frac{2\pi}{\lambda}
{\mathbf{t}}_n^{\mathsf{T}}{\bm\rho}_{\rm{b}}^{\ell,\ell'}+\theta_{\rm{b}}^{\ell,\ell'}){\bm\rho}_{\rm{b}}^{\ell,\ell'}}{-\frac{\lambda L_{\rm{b}}}{\lvert\sigma_{{\rm{b}},\ell}\sigma_{{\rm{b}},\ell'}\rvert}}.
\end{split}
\end{equation}
}The algorithm for optimizing $\mathbf{T}$ is presented in Algorithm \ref{Algorithm1}. Convergence is guaranteed due to the upper-bounded nature of the channel power gain. The complexity of Algorithm \ref{Algorithm1} scales as ${\mathcal{O}}(I_{\rm{gd}}NL_{\rm{b}}^2\log_2\frac{1}{u_{\min}})$, where $I_{\rm{gd}}$ is the number of iterations, and $u_{\min}$ represents the desired accuracy.

\begin{algorithm}[!t]
\algsetup{linenosize=\tiny} \scriptsize
  \caption{Gradient-Based Algorithm for Optimizing $\mathbf{T}$}
  \label{Algorithm1}
  \begin{algorithmic}[1]
    \STATE Initialize ${\mathbf{T}}^{0}=[{\mathbf{t}}_1^0 \ldots {\mathbf{t}}_N^0]$, the maximum iteration number $I_{\rm{gd}}$, step size $u_{\rm{ini}}$, the minimum tolerance step size $u_{\min}$, and set the current iteration $a=0$;
    \REPEAT
    \FORALL{$n=1:N$} 
    \STATE Compute the gradient $\nabla_{{\mathbf{t}}_n^{a}}{f_n}$ and set $u=u_{\rm{ini}}$;
      \REPEAT
      \STATE Compute $\hat{\mathbf{t}}_n={\mathbf{t}}_n^{a}+u\cdot\nabla_{{\mathbf{t}}_n^{a}}{f_n}$ and set $u=u/2$;     
      \UNTIL{$\hat{\mathbf{t}}_n\in{\mathcal{S}}_n \& f_n(\hat{\mathbf{t}}_n)>f_n({\mathbf{t}}_n^{a})$ or $u<u_{\min}$};
      \STATE Set ${\mathbf{t}}_n^{a}=\hat{\mathbf{t}}_n$ and update ${\mathbf{t}}_n^{a+1}=\hat{\mathbf{t}}_n$;
    \ENDFOR
      \STATE Update $a=a+1$;
    \UNTIL{convergence or $a>I_{\rm{gd}}$}.
  \end{algorithmic}
\end{algorithm} 

\subsection{Jamming Design}
After solving \eqref{Power_Min_Problem2}, the remaining power $P-P_{\rm{T}}^{\star}$ is allocated for generating AN signals within the null space of ${\mathbf{h}}_{\rm{b}}$. Given the complete unavailability of the CSI of ${\mathbf{h}}_{\rm{e}}$, it becomes infeasible to optimize the transmit covariance ${\mathbf{C}}={\mathbbmss{E}}\{{\mathbf{z}}{\mathbf{z}}^{\mathsf{H}}\}$ to minimize the leakage rate. Consequently, we employ isotropic signaling with equal power allocation to transmit AN signals across each dimension of the null space of ${\mathbf{h}}_{\rm{b}}$. Given that the dimension of this null space is $N-1$, the transmit covariance for AN can be expressed as follows:
{\setlength\abovedisplayskip{2pt}
\setlength\belowdisplayskip{2pt}
\begin{equation}\label{AN_Matrix}
{\mathbf{C}}=\frac{P-P_{\rm{T}}^{\star}}{N-1}{\mathbf{U}}_{\rm{b}}{\mathbf{U}}_{\rm{b}}^{\mathsf{H}},
\end{equation}
}where the columns of the semi-unitary matrix ${\mathbf{U}}_{\rm{b}}\in{\mathbbmss{C}}^{N\times (N-1)}$ constitute all $N-1$ eigenvectors corresponding to zero eigenvalues of ${\mathbf{h}}_{\rm{b}}{\mathbf{h}}_{\rm{b}}^{\mathsf{H}}$. Using the completeness of the basis formed by ${\mathbf{h}}_{\rm{b}}$ and the columns of ${\mathbf{U}}_{\rm{b}}$, we obtain
{\setlength\abovedisplayskip{2pt}
\setlength\belowdisplayskip{2pt}
\begin{align}
{\mathbf{U}}_{\rm{b}}{\mathbf{U}}_{\rm{b}}^{\mathsf{H}}={\mathbf{I}}_N-\frac{{\mathbf{h}}_{\rm{b}}}{\lVert{\mathbf{h}}_{\rm{b}}\rVert}
\frac{{\mathbf{h}}_{\rm{b}}^{\mathsf{H}}}{\lVert{\mathbf{h}}_{\rm{b}}\rVert}={\mathbf{I}}_N-\frac{1}{\lVert{\mathbf{h}}_{\rm{b}}\rVert^2}{\mathbf{h}}_{\rm{b}}{\mathbf{h}}_{\rm{b}}^{\mathsf{H}}.
\end{align}
}\subsection{Further Discussions}
\subsubsection{Secrecy Rate}
Based on our previous derivations, the secrecy rate can be expressed as follows \cite{Khisti2010}:
{\setlength\abovedisplayskip{2pt}
\setlength\belowdisplayskip{2pt}
\begin{align}\label{Secrecy_Rate_1}
{\mathcal{R}}_{\rm{s}}&=\log_2\left(1+\frac{\lvert{\mathbf{h}}_{\rm{b}}^{\mathsf{H}}{\mathbf{w}}^{\star}\rvert^2}{\sigma_{\rm{b}}^2}\right)
-\log_2\left(1+\frac{\lvert{\mathbf{h}}_{\rm{e}}^{\mathsf{H}}{\mathbf{w}}^{\star}\rvert^2}{\sigma_{\rm{e}}^2+{\mathbf{h}}_{\rm{e}}^{\mathsf{H}}{\mathbf{C}}{\mathbf{h}}_{\rm{e}}}\right).
\end{align}
}Inserting ${\mathbf{w}}^{\star}=\sqrt{P_{\rm{T}}}\frac{{\mathbf{h}}_{\rm{b}}}{\lVert{\mathbf{h}}_{\rm{b}}\rVert}$ and \eqref{AN_Matrix} into \eqref{Secrecy_Rate_1} gives
{\setlength\abovedisplayskip{2pt}
\setlength\belowdisplayskip{2pt}
\begin{equation}\label{Secrecy_Rate_2}
\begin{split}
{\mathcal{R}}_{\rm{s}}&=\log_2\left(1+\frac{P_{\rm{T}}\lVert{\mathbf{h}}_{\rm{b}}\rVert^2}{\sigma_{\rm{b}}^2}\right)\\
&-\log_2\left(1+\frac{\frac{P_{\rm{T}}}{\lVert{\mathbf{h}}_{\rm{b}}\rVert^2}\lvert{\mathbf{h}}_{\rm{b}}^{\mathsf{H}}{\mathbf{h}}_{\rm{e}}\rvert^2}
{\sigma_{\rm{e}}^2+\frac{P-P_{\rm{T}}}{N-1}(\lVert{\mathbf{h}}_{\rm{e}}\rVert^2-\frac{\lvert{\mathbf{h}}_{\rm{b}}^{\mathsf{H}}{\mathbf{h}}_{\rm{e}}\rvert^2}{\lVert{\mathbf{h}}_{\rm{b}}\rVert^2})}\right).
\end{split}
\end{equation}
}Given that the target QoS SNR of Bob is satisfied, we have $P_{\rm{T}}=\frac{\gamma{\sigma_{\rm{b}}^2}}{\lVert{\mathbf{h}}_{\rm{b}}\rVert^2}$. Furthermore, assuming $\sigma_{\rm{b}}^2=\sigma_{\rm{e}}^2$, and substituting it into \eqref{Secrecy_Rate_2}, we obtain
{\setlength\abovedisplayskip{2pt}
\setlength\belowdisplayskip{2pt}
\begin{equation}
\begin{split}
{\mathcal{R}}_{\rm{s}}=\log_2\left(1+\gamma\right)-\log_2\left(1+\frac{\gamma\rho{\lVert{\mathbf{h}}_{\rm{e}}\rVert^2}/{\lVert{\mathbf{h}}_{\rm{b}}\rVert^2}}
{1+\frac{(P-P_{\rm{T}})\lVert{\mathbf{h}}_{\rm{e}}\rVert^2(1-\rho)}{(N-1)\sigma_{\rm{e}}^2}}\right),\nonumber
\end{split}
\end{equation}
}where $\rho=\frac{\lvert{\mathbf{h}}_{\rm{b}}^{\mathsf{H}}{\mathbf{h}}_{\rm{e}}\rvert^2}{\lVert{\mathbf{h}}_{\rm{b}}\rVert^2\lVert{\mathbf{h}}_{\rm{e}}\rVert^2}$ represents the channel correlation factor. It is important to note that the secrecy rate is a function of the location matrix $\mathbf{T}$.
\subsubsection{Linear Array}
\begin{figure}[!t]
\centering
    \subfigbottomskip=-5pt
	\subfigcapskip=-5pt
\setlength{\abovecaptionskip}{0pt}
\includegraphics[height=0.20\textwidth]{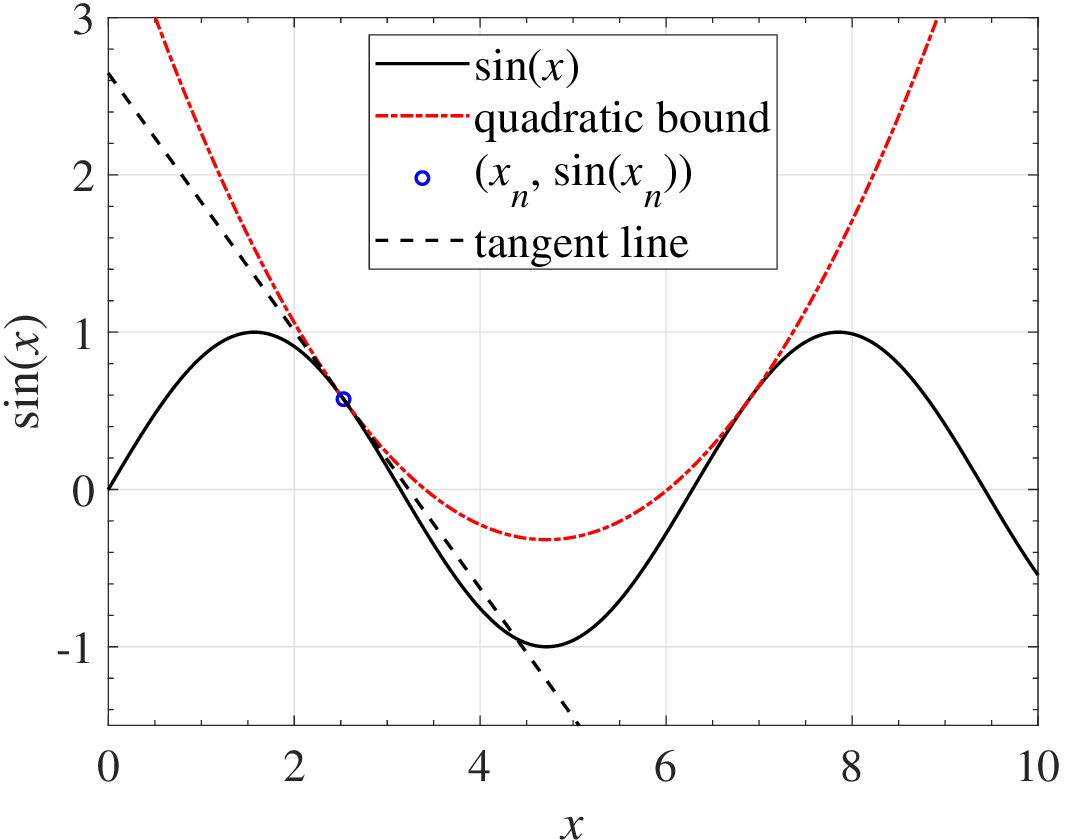}
\caption{Approximating the $\sin(\cdot)$ function by a convex quadratic function.}
\label{Figure:BSUM}
\vspace{-5pt}
\end{figure}

Next, we consider a special case where Alice employs a linear array along the $x$-axis. In this scenario, the position of the $n$th MA is given by ${\mathbf{t}}_n=[x_n,0]^{\mathsf{T}}$ for $n\in{\mathcal{N}}$, meaning each MA can only move along one dimension. Thus, Bob's channel power can be simplified as follows:
{\setlength\abovedisplayskip{2pt}
\setlength\belowdisplayskip{2pt}
\begin{align}\label{Channel_Power_Linear_Array}
\lVert{\mathbf{h}}_{\rm{b}}\rVert^2=
\frac{\mu_{\rm{b}}}{L_{\rm{b}}}\sum_{n=1}^{N}\left\lvert\sum_{\ell=1}^{L_{\rm{b}}}\sigma_{{\rm{b}},\ell}{\rm{e}}^{-{\rm{j}}\frac{2\pi}{\lambda}
x_n\sin{\theta_{{\rm{b}},\ell}}\cos{\phi_{{\rm{b}},\ell}}}\right\rvert^2.
\end{align}
}Given the objective function in \eqref{Channel_Power_Linear_Array}, the design of the MAs' positions $\{{x}_n\}_{n=1}^{N}$ can be achieved using the block successive upper-bound minimization (BSUM) method \cite{Razaviyayn2013}. Specifically, given $\{{x}_{n'}\}_{n'\ne n}$, the design of $x_n$ can be formulated as follows:
{\setlength\abovedisplayskip{2pt}
\setlength\belowdisplayskip{2pt}
\begin{subequations}\label{Power_Min_Problem_Instantaneous_Linear}
\begin{align}
\min_{x_n}~&f_n(x_n)\triangleq \left\lvert\sum\nolimits_{\ell=1}^{L_{\rm{b}}}\sigma_{{\rm{b}},\ell}{\rm{e}}^{-{\rm{j}}\frac{2\pi}{\lambda}
x_n\sin{\theta_{{\rm{b}},\ell}}\cos{\phi_{{\rm{b}},\ell}}}\right\rvert^2\label{Power_Min_Problem_Instantaneous_Linear_Obj}\\
{\rm{s.t.}}~&x_n\in[-A/2,A/2],\lvert x_n-x_{n'}\rvert\geq D,n\ne n'.\label{Power_Min_Problem_Instantaneous_Linear_Cons2}
\end{align}
\end{subequations}
}The objective function can be further simplified by omiting the constant term as follows:
{\setlength\abovedisplayskip{2pt}
\setlength\belowdisplayskip{2pt}
\begin{align}
f_n(x_n)=
\sum\nolimits_{\ell=1}^{L_{\rm{b}}}\sum\nolimits_{\ell'\ne \ell}\lvert\sigma_{{\rm{b}},\ell}\sigma_{{\rm{b}},\ell'}\rvert
\sin(x_n{{\rho}_{\rm{b}}^{\ell,\ell'}}+{\theta_{\rm{b}}^{\ell,\ell'}}),\nonumber
\end{align}
}where ${{\rho}_{\rm{b}}^{\ell,\ell'}}=\frac{2\pi}{\lambda}(\sin{\theta_{{\rm{b}},\ell}}\cos{\phi_{{\rm{b}},\ell}}-\sin{\theta_{{\rm{b}},\ell'}}\cos{\phi_{{\rm{b}},\ell'}})$. By leveraging the properties of sine functions, a quadratic upper bound for $f_n(x_n)$ can be established. For a given $x_n$, the quadratic upper bound of $\sin(x{{\rho}_{\rm{b}}^{\ell,\ell'}}+{\theta_{\rm{b}}^{\ell,\ell'}})$ is denoted by $a_{\rm{b}}^{\ell,\ell'}(x-b_{\rm{b}}^{\ell,\ell'})^2+c_{\rm{b}}^{\ell,\ell'}$, where $\{a_{\rm{b}}^{\ell,\ell'},b_{\rm{b}}^{\ell,\ell'},c_{\rm{b}}^{\ell,\ell'}\}$ are determined by the tangent line of $\sin(x{{\rho}_{\rm{b}}^{\ell,\ell'}}+{\theta_{\rm{b}}^{\ell,\ell'}})$ at $x=x_n$ and the value of $x_n{{\rho}_{\rm{b}}^{\ell,\ell'}}+{\theta_{\rm{b}}^{\ell,\ell'}}$. A simple example of this is illustrated in {\figurename} {\ref{Figure:BSUM}}. The quadratic upper bound of $f_n(x)$ at $x_n$ is thus:
{\setlength\abovedisplayskip{2pt}
\setlength\belowdisplayskip{2pt}
\begin{align}
f_n(x)\leq\sum\nolimits_{\ell=1}^{L_{\rm{b}}}\sum\nolimits_{\ell'\ne \ell}\lvert\sigma_{{\rm{b}},\ell}\sigma_{{\rm{b}},\ell'}\rvert
(a_{\rm{b}}^{\ell,\ell'}(x-b_{\rm{b}}^{\ell,\ell'})^2+c_{\rm{b}}^{\ell,\ell'}).\nonumber
\end{align}
}By iteratively minimizing this upper bound for different $x_n$, i.e., solving the following subproblem:
{\setlength\abovedisplayskip{2pt}
\setlength\belowdisplayskip{2pt}
\begin{subequations}
\begin{align}
\min_{x_n}~&\sum_{\ell=1}^{L_{\rm{b}}}\sum_{\ell'\ne \ell}\lvert\sigma_{{\rm{b}},\ell}\sigma_{{\rm{b}},\ell'}\rvert
(a_{\rm{b}}^{\ell,\ell'}(x-b_{\rm{b}}^{\ell,\ell'})^2+c_{\rm{b}}^{\ell,\ell'})\\
{\rm{s.t.}}~&x_n\in[-A/2,A/2],\lvert x_n-x_{n'}\rvert\geq D,n\ne n',\label{Power_Min_Problem_Instantaneous_Linear_Cons2}
\end{align}
\end{subequations}
}a stationary-point solution to the channel power maximization problem can be obtained \cite{Razaviyayn2013}. Since the minimization of a quadratic function has a closed-form solution, each BSUM iteration involves low computational complexity. Due to space limitations, further details are omitted and will be presented in a full version of this work.

\section{Numerical Results}\label{Simulations}
In this section, computer simulations are conducted to validate the effectiveness of our proposed transmission scheme in improving the secrecy performance of the MA-aided secure communication system. Unless otherwise specified, we set $N=4$, $D = \frac{\lambda}{2}$, $u_{\min}=10^{-3}$, $u_{\rm{ini}}=10$, $I_{\rm{gd}}=30$, $\mu_i=-110$ dBm for $i\in\{{\rm{b}},{\rm{e}}\}$, and $\frac{P}{\sigma_{\rm{b}}^2}=10$. The noise power is given by $\sigma_i^2=N_0B$ ($\forall i$) with a bandwidth of $B = 1$ MHz and an effective noise power density of $N_0=-174$ dBm/Hz. For the channel model, we assume $L_i=4$ ($\forall i$). The elevation and azimuth angles are randomly selected within the range $[0, \pi]$. We compare the performance of our proposed algorithms with an FPA-based benchmark scheme, where Alice is equipped with an FPA-based uniform linear array consisting of $N$ antennas spaced by $\frac{\lambda}{2}$. The presented numerical results are averaged over $1000$ independent channel realizations with randomly initialized optimization variables.

\begin{figure}[!t]
\centering
    \subfigbottomskip=-5pt
	\subfigcapskip=-5pt
\setlength{\abovecaptionskip}{0pt}
\includegraphics[height=0.20\textwidth]{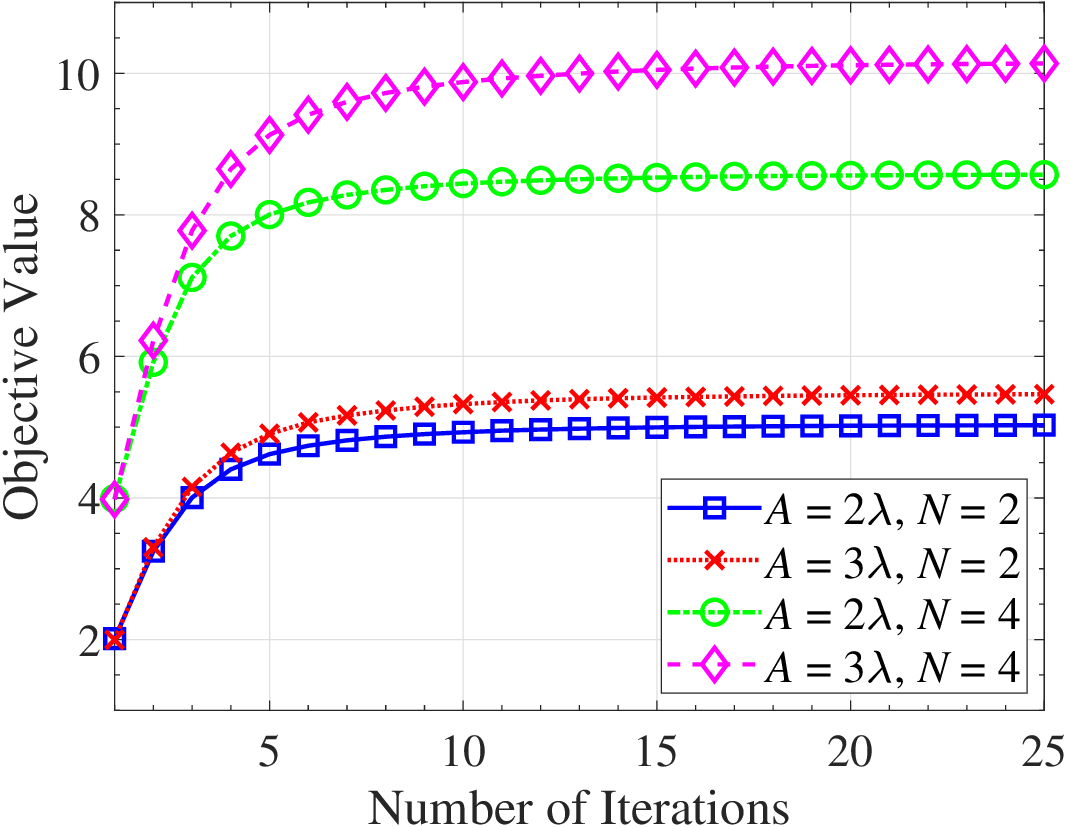}
\caption{Convergence of the proposed algorithm.}
\label{fig1}
\vspace{-5pt}
\end{figure}

\begin{figure}[!t]
    \centering
    \subfigbottomskip=0pt
	\subfigcapskip=-5pt
\setlength{\abovecaptionskip}{0pt}
    \subfigure[Secrecy rate vs. the target SNR.]
    {
        \includegraphics[height=0.18\textwidth]{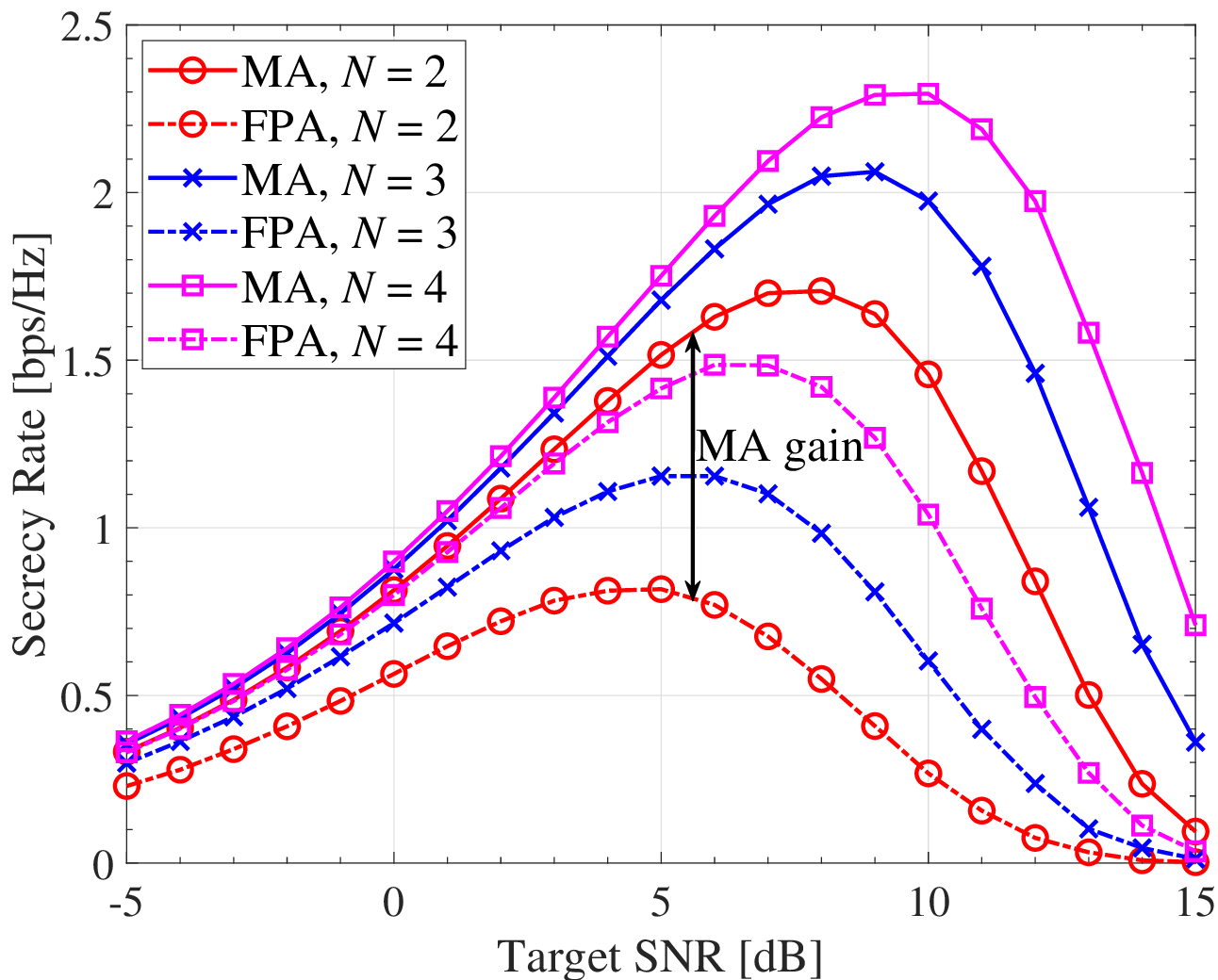}
	   \label{fig2a}	
    }
   \subfigure[Secrecy rate vs. the region size.]
    {
        \includegraphics[height=0.18\textwidth]{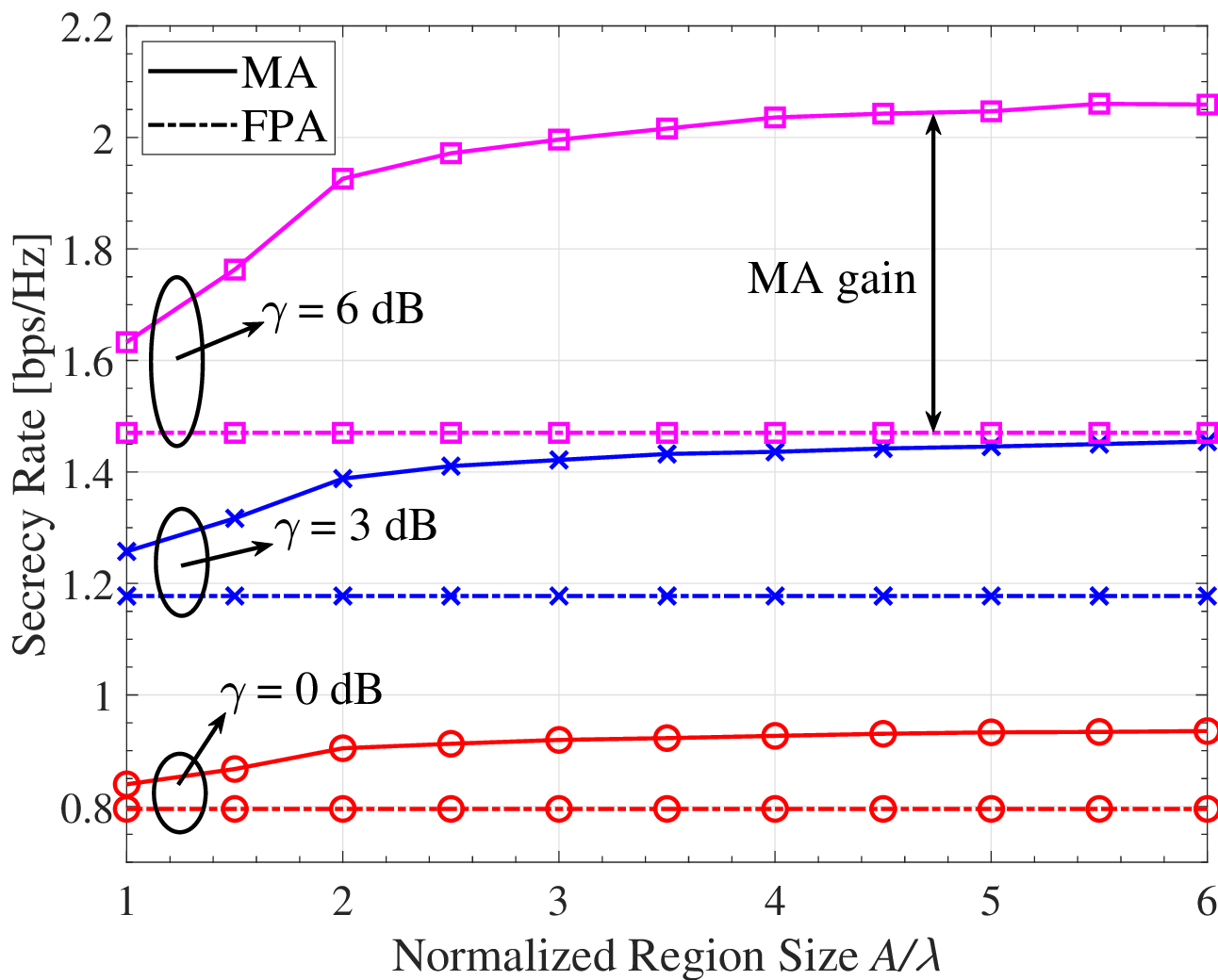}
	   \label{fig2b}	
    }
\caption{Secrecy rate performance of the proposed scheme.}
    \label{Figure2}
    \vspace{-10pt}
\end{figure}

In {\figurename} {\ref{fig1}}, we illustrate the convergence performance of our proposed gradient-based algorithm for optimizing the antenna position matrix $\mathbf{T}$. As shown in {\figurename} {\ref{fig1}}, Bob's channel power $\lVert{\mathbf{h}}_{\rm{b}}\rVert^2$ (the objective value) increases rapidly with the number of iterations. Specifically, the gradient-based algorithm converges within approximately $25$ iterations under different system setups. This demonstrates the effectiveness of the gradient-based algorithm in optimizing MAs' positions.

In {\figurename} {\ref{fig2a}}, we present the secrecy rate of the proposed and benchmark schemes versus Bob's SNR target $\gamma$. It is observed that, for the same SNR target, our proposed scheme achieves a higher secrecy rate than the FPA-based schemes. This is because MAs provide additional spatial DoFs in terms of antenna positioning, allowing them to achieve the same SNR target with less power than FPAs. Consequently, more residual power can be allocated to AN signaling to jam Eve, resulting in a higher secrecy rate. Furthermore, as shown in the graph, increasing $\gamma$ significantly enhances the secrecy rate because the legitimate information rate is greatly increased. However, when $\gamma$ is set too high, insufficient power remains for AN signaling, causing the information leakage rate to dominate, and the achieved secrecy rate decreases. If $\gamma$ exceeds the total available power, making the problem \eqref{Power_Min_Problem_Instantaneous} infeasible, transmission fails. In {\figurename} {\ref{fig2b}}, we plot the secrecy rate versus the normalized region size $A/\lambda$. It is observed that the proposed MA-based schemes outperform FPA systems in terms of secrecy rate, and the performance gain increases with the region size. Additionally, the MA-based scheme exhibits convergence when the value of $A/\lambda$ exceeds $4$, suggesting that optimal secrecy performance for MA-aided secure communication systems can be attained within a finite region.
\section{Conclusion}
In this paper, we proposed an MA-aided secure transmission framework without relying on the eavesdropper's CSI. To enhance the secrecy rate, we introduce a joint beamforming and jamming scheme, which optimizes the MAs' positions to improve transmit beamforming and AN signaling. Simulation results have validated the effectiveness of the proposed algorithms, demonstrating that the MA-based scheme outperforms FPA-based approaches in terms of improving the secrecy rate.

\end{document}